\documentclass[prb,twocolumn,showpacs,amsmath,amssymb]{revtex4}
\usepackage{graphicx}
\usepackage{tabularx}
\usepackage{dcolumn}
\usepackage{bm}
\usepackage{color}

\begin{document}
\title{
Interplay of spin magnetism, orbital magnetism, and atomic structure in 
layered van der Waals ferromagnet VI$_3$}

\author{L. M. Sandratskii, K. Carva}
\affiliation{Faculty of Mathematics and Physics, Charles University, 12116 Prague,
Czech Republic}

\begin{abstract}
Recently discovered ferromagnetism of the layered van der Waals material VI$_3$ attracts much 
research attention. Despite substantial progress, 
in the following important aspects no consensus has been reached:
(i) a possible deviation of the easy axis from
the normal to the VI$_3$ layers, (ii) a possible inequivalence of the V atoms, (iii) the value of the V 
magnetic moments. The theoretical works differ in the conclusions on the conduction nature of the system, 
the value and the role of the V orbital moments. To the best of our knowledge there is no 
theoretical works addressing issues (i) and (ii) and only one work dealing with the reduced
value of the V moment. 
By combining the symmetry arguments with density functional theory (DFT) and DFT+$U$ 
calculations we have shown
that the antidimerization distortion of the crystal structure 
reported by Son {\it et al}. [Phys. Rev. B {\bf 99}, 041402(R) (2019)] 
must lead to the deviation of the 
easy axis from the normal to the VI$_3$ layers in close correlation with the experimental results.
The antidimerization accompanied by 
the breaking 
the inversion symmetry leads to the inequivalence of the V atoms.
Our DFT+$U$ calculations result in large value $\sim$0.8~$\mu_B$ of the V orbital moments 
of the V atoms
leading to reduced total V moment in agreement with a number of experimental results
and with the physical picture suggested by Yang {\it et al}. [Phys. Rev. B {\bf 101}, 100402(R) (2020)]. 
We obtained large 
intraatomic noncollinearity of the V spin and orbital moments revealing strong competition   
between 
effects coursed by the on-site electron correlation, spin-orbit coupling, and 
interatomic hybridization
\textcolor{black}{since pure intraatomic effects lead to collinear spin and orbital
moments.
}
Our calculations confirm the experimental results of strong magnetoelastic 
coupling revealing itself in the strong dependence of the magnetic properties 
on the distortion of the atomic structure. 
\end{abstract}

\maketitle

\section{Introduction}

The search for new magnetic 2D materials for spintronic applications is one of the hot
topics of the present-day solid state physics (see, e.g., Ref.~\onlinecite{Gibrtini2019} for a review).
In recent few years, much research attention has been attracted to the layered van der Waals (vdW) 
trihalide VI$_3$ \cite{He2016,Son2019,Dolezal2019,An2019,Kong2019,Tian2019,Yan2019,Gati2019,
Jia2020,Yang2020,Subhan2020,Wang2020,Liu2020,Huang2020}.
The conclusions of different studies of VI$_3$ 
appeared to contain both important agreements and important differences.
Thus the experimental works agree 
on the insulating nature of the system, and ferromagnetic type of the magnetic 
structure \cite{Tian2019,Yan2019,Son2019,Kong2019,Liu2020}. 
In most 
studies the $z$ axis, orthogonal to the VI$_3$ layers, is treated as
the easy axis, and all V atoms are considered to be equivalent. 
On the other hand, there are experimental reports pointing to significant deviation of the easy axis 
from the $z$ axis (Ref.~\onlinecite{Yan2019}, supplementary material to Ref.~\onlinecite{Dolezal2019}) 
and the inequivalence of the V atoms \cite{Gati2019}. Remarkably, there is no 
consensus also on the value of the V magnetic moments: the reported experimental estimations 
differ strongly. Taking as an example the moments obtained in the measurements in magnetic field 
parallel to the $z$ axis we find the following values (all in~$\mu_B$):  2.47 (Ref.~\onlinecite{Tian2019}),
$\sim$2 (Ref.~\onlinecite{Yan2019}), 
1.3 (Ref.~\onlinecite{Son2019}), 1.2 (Ref.~\onlinecite{Kong2019}), 0.95 (Ref.~\onlinecite{Liu2020}).
 
On the theoretical side, all publications consider the ferromagnetic state with the easy $z$ axis 
and equivalent V atoms
\cite{He2016,An2019,Jia2020,Yang2020}. 
All studies find the V spin moment
close to 2.0~$\mu_B$. Some 
works 
\cite{He2016,An2019}
compare this value with experimental magnetic moment claiming good agreement. 
The orbital moment (OM) is usually not reported. An important exception is 
a recent work by Yang {\it et al.} \cite{Yang2020} arguing that in the ground state 
the V atoms have a large OM of about 1~$\mu_B$. 
This magnetic state was obtained in the DFT+$U$ \cite{Anisimov1997} 
calculation taking into account the correlation of the V 3d electrons.
The large OM
opposite to the spin moment leads to the reduced value of the total atomic moment.

Another important difference in the conclusions of the theoretical studies 
concerns the conduction nature of the system. 
Several works (Refs.~\onlinecite{He2016,Jia2020}) report the half-metallic 
electronic structure and find 
Dirac or Weyl states close to the Fermi energy in the conducting spin-up subsystem. 
\textcolor{black}{
Other works obtain Mott insulator state in agreement with experiment \cite{An2019,Yang2020,Wang2020}. 
}
In Ref.~\onlinecite{Wang2020} the authors discuss
the possibility of the convergence of the DFT+$U$ calculations to different states 
that can explain difference of the results obtained within
apparently similar theoretical approaches. 

The goal of the present paper is to contribute to the understanding of VI$_3$ in the 
following aspects.
First, we address the findings of the experimental works that, despite their
importance, have not yet been treated theoretically. 
As mentioned above, these findings are the deviation of the easy axis
from the normal to the VI$_3$ layers and the inequivalence of the V atoms.  
Another characteristic feature of our work
is a systematic attention paid to the properties of the atomic OMs.
In general, the OMs play important role in the magnetism of various types of systems. 
They provide insight into the physical consequences of spin-orbit coupling (SOC). 
In particular, the properties of the OMs are
known to be closely related with the properties of the magnetic anisotropy 
(see, e.g., Refs.~\onlinecite{Bruno1995,Sandratskii13,Soares2014,Sandratskii15}).
In the case of VI$_3$ the attention to the OMs is additionally stimulated
by the experimental reports of reduced values of the atomic moments and by the 
physical picture suggested in Ref.~\onlinecite{Yang2020} whose crucial feature is a large OM
of the V atoms.
We also contribute to the discussion of the multiple convergence of the calculations 
taking into account electron correlations within the framework of the DFT+$U$ approach.

Our theoretical methodology consists in the combination of the symmetry-based analysis 
of magnetic states with the 
DFT and DFT+$U$ calculations. The strong feature of the symmetry treatment is a wide
independence of the drawn conclusions from the details of the 
approach used in the calculation of the electronic properties.
On the other hand, the symmetry analysis does not provide quantitative 
estimates of the physical quantities. Therefore, we perform the DFT based
calculations to evaluate the magnitudes of the symmetry-predicted effects. 

The paper is organized as follows. In Sec.~\ref{Sec_Method of calculations} we 
describe the method of calculations. Section~\ref{Sec_Lattice} presents the atomic structure
and discusses the symmetry aspects of the study. Section~\ref{Sec_Results} contains the results
of the calculations and their discussion. Section~\ref{Sec_conclus} is devoted to the conclusions.

\section{Method of calculations}
\label{Sec_Method of calculations}

The calculations are performed with the augmented spherical waves (ASW)
method \cite{Williams1979,Eyert2012}
generalized to deal with noncollinear magnetism and spin-orbit coupling
\cite{Sandratskii1998}.  The generalized gradient approximation (GGA) to the
energy functional \cite{Perdew1996} is employed in the calculations.
The DFT+$U$ method in the form suggested by Dudarev {\it et al.} \cite{Dudarev1998}
was applied to examine the influence of the on-site correlation of the 3d electrons
on the magnetic moments and energies of magnetic configurations.
\textcolor{black}{
We used the $U_{eff}$ parameter equal to 0.2~Ry ($\sim$2.72~eV)
what is a reasonable value for this material \cite{An2019,Yang2020}.
}
The most of calculations
were performed with {\bf k}-mesh 12$\times$12$\times$12. 

An important quantity of the DFT+$U$
approach is the orbital density matrix $n$ of the correlated atomic states. 
It enters the method
with the prefactor $U$ leading to the orbital dependence of the electron potential \cite{Dudarev1998}
\begin{equation}
V_{m,m'}=-U(n_{m,m'}-\frac{1}{2}\delta_{m,m'}).
\label{Eq_Vmm}
\end{equation}
\textcolor{black}{
In the paper we work in the basis of complex spherical harmonics $Y_{lm}$.
}
We consider the correlation of the V 3d electrons. Therefore the orbital quantum number $l$ is equal to 2
and the orbital dependence of the potential is given by the dependence on the magnetic quantum number $m$.
All other indices characterizing orbitals are omitted.
The diagonal elements $n_{m,m}$ of the orbital density matrix give the occupations of the corresponding
$m$ orbitals. Examples of the implementation of the  $n$ matrix calculation within the DFT
methods can be found in Refs.~\onlinecite{Shick1999,Bengone2000}.
 
\textcolor{black}{
The operator of the spin-orbit coupling is taken in the form \cite{Sandratskii13}
\begin{multline}
\label{Hso}
{\bf H}_{so}=\frac{1}{(2c)^2}\frac{1}{r}
     \bigg [
     \left(\begin{array}{cc}
     \frac{1}{M_{+}^2}\frac{dV^{+}}{dr} & 0\\
      0&\frac{1}{M_{-}^2}
      \frac{dV^{-}}{dr}
       \end{array}\right)
\sigma_z\hat{l}_z \\
 + \frac{1}{M_{av}^2}\frac{dV^{av}}
{dr}
   (\sigma_x\hat{l}_x+\sigma_y\hat{l}_y)
     \bigg ]
\end{multline}
where $V^{+}$ and $V^{-}$ are spin-up and spin-down electron potentials,
 \begin{equation}
V^{av}=\frac{1}{2}
(V^{+}+V^{-})
    \label{so2}
 \end{equation}
and
 \begin{equation}
M_\alpha=\frac{1}{2}(1-\frac{1}{c^2}V^\alpha)\quad ,\alpha=av
,+,-.
 \label{so3}
 \end{equation}
$\sigma_x,\sigma_y,\sigma_z$ are the Pauli matrices and $\hat{l}_x,\hat{l}_y,\hat{l}_z$ are
the operators of the components of the orbital momentum,
$r$ is the distance from the center of atomic surface, $c$ is
the light velocity.  
The SOC is taken into account for both V and I atoms.
}

\textcolor{black}{
We calculate the vectors of spin $\mathbf{m}_s^\nu$ and orbital $\mathbf{m}_{o}^\nu$ moments 
of the $\nu$th atom as
\begin{equation}
\mathbf{m}_s^\nu=\sum_{\mathbf{k}n}^{occ}\int_{\Omega_{\nu}}\psi_{\mathbf{k}n}^\dag\mathbf{\sigma}
\psi_{\mathbf{k}n}d\mathbf{r}
\end{equation}
\begin{equation}
\mathbf{m}_{o}^\nu=\sum_{\mathbf{k}n}^{occ}\int_{\Omega_{\nu}}\psi_{\mathbf{k}n}^\dag\mathbf{\hat{l}}
\psi_{\mathbf{k}n}d\mathbf{r}
\end{equation}
where $\mathbf{\sigma}=(\sigma_x,\sigma_y,\sigma_z)$
and $\mathbf{\hat{l}}=(\hat{l}_x,\hat{l}_y,\hat{l}_z)$, $\psi_{\mathbf{k}n}$ is
the wave function of the Kohn-Sham state corresponding to wave vector $\mathbf{k}$
and band index $n$. The sum is taken over occupied states.
The integrals are carried out over $\nu$th atomic sphere. 
}

Due to the orbital-dependent potential term (Eq.~\ref{Eq_Vmm})
the occupied orbitals tend to lower their energies 
whereas empty orbitals tend to increase their energies.
This feature makes the DFT+$U$ approach an adequate tool for the study of 
the enhancement of the orbital magnetic moment \cite{Solovyev1998}.
It is well known that standard DFT calculations underestimate the value of the OM.
There is a similarity in the physical mechanisms of the OM enhancement
within the DFT+$U$ method \cite{Solovyev1998} and the method of the orbital polarization correction
by Eriksson {\it et al}. \cite{Eriksson1990}
The latter approach is based on the simulation of the intraatomic second Hund's rule in the
DFT calculations. There is, however, an important difference in the scale of the
terms responsible for the orbital polarization enhancement. 
In the DFT+$U$ scheme the scale is governed by the Hubbard 
parameter $U$ \cite{Solovyev1998,Dudarev1998}
whereas in the second Hund's rule this is a usually smaller parameter often referred to as $B$.  

The hybridization of the correlated states with the states of ligands produces 
physical mechanism counteracting the energy shifts of the correlated states.
This results in a complex interplay of intraatomic and interatomic interactions
that is taken into account in the DFT+$U$ calculations. 

\section{Crystal lattice and symmetry aspects}
\label{Sec_Lattice}

\subsection{Undistorted $R\overline{3}$ atomic structure}

\begin{figure}[t]
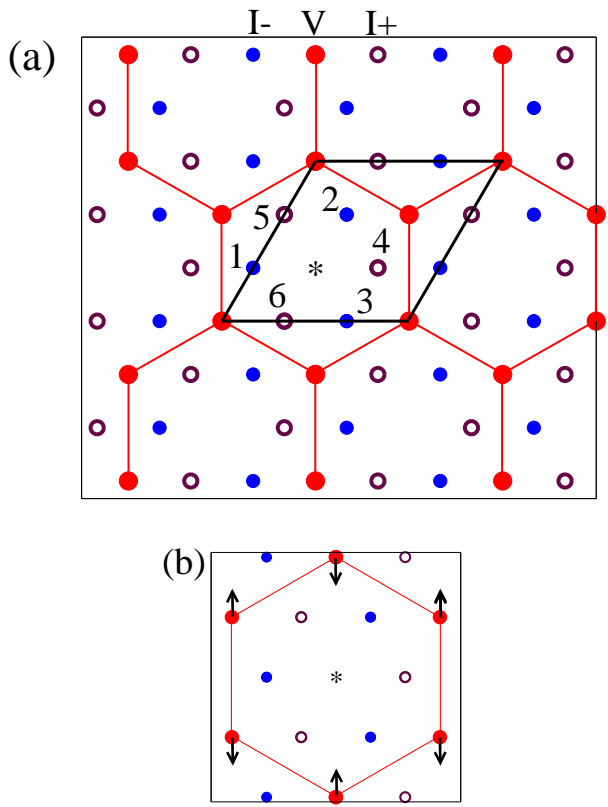

\includegraphics*[width=8cm]{FIG_lattice.eps}
\vspace{0.5cm}\\
\includegraphics*[width=4cm]{FIG_AD.eps}
\caption{(a) Projection of the VI$_3$ monolayer on the $xy$ plane.
The V atoms build the honeycomb lattice. The I atoms labeled as 
I- (I+) form the layers below (above) the V layer. The rhombus in the 
center of the figure gives the in-plane unit cell. For convenience of 
reference the six I atoms in the unit cell are numbered. The asterisk
shows the position of the inversion center.
(b) The arrows show the directions of the antidimerization shifts of 
the V atoms. In the calculations the value of the shift was chosen 
to give the 2\% reduction of the distance between atoms moving
in the figure towards each other.
}
\label{Fig_lat}
\end{figure}
\begin{figure}[t]
\includegraphics*[width=8cm]{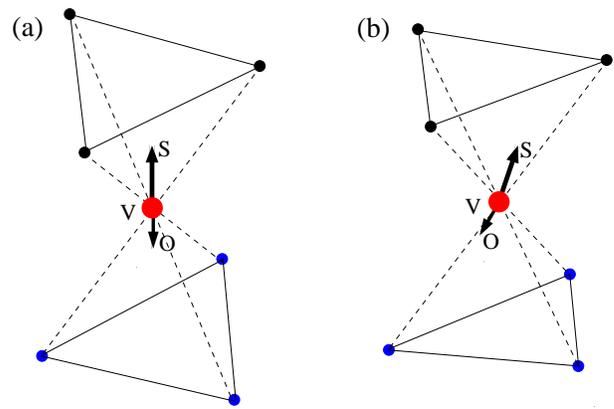}
\textcolor{black}{
\caption{Schematic figure illustrating deviation of the V atomic moments from the $z$ axis
in the distorted atomic structure. (a) $R\overline{3}$ atomic structure. Both spin (S) and orbital (O) moments
of the V atoms are collinear to the $z$ axis. (b) In the distorted atomic structure the spin and orbital
moments deviate from the $z$ axis by different angles becoming noncollinear to each other.
}
}
\label{Fig_mod}
\end{figure}
The system experiences structural phase transition at $\sim$80~K  that is above the temperature
of the magnetic phase transition. 
One VI$_3$ layer consists of the honeycomb V lattice and two I layers lying above and below the 
V layer (Fig.~\ref{Fig_lat}a).
The neighboring VI$_3$ layers are weakly bounded through the vdW interaction. The stacking of the 
layers we consider is ABC\cite{Kong2019}. The structural phase transition leads to the distortion of the honeycomb 
lattice. Son {\it et al}. \cite{Son2019} describe this distortion in terms of antidimerization (AD) of the V atoms. 
We performed calculations for both $R\overline{3}$ and distorted (Fig.~\ref{Fig_lat}b) atomic lattices. 
In this section we discuss the 
symmetries of the lattices and their physical consequences.

We begin with the symmetry properties of VI$_3$ in the $R\overline{3}$ crystal lattice. 
The $R\overline{3}$ crystal structure has 6 symmetry operations. 
The two generaters of the symmetry group are
120$^\circ$-rotation about the $z$ axis, $C_{3z}$, and space inversion. 

The unit cell contains two formula units of VI$_3$.
The equivalence of the atoms belonging to one unit cell  
does not follow from the translational invariance of the lattice
and, if present, must be the consequence of the point symmetry operations. 
The $C_{3z}$ axes go through the positions of the V atoms leaving them invariant 
under $C_{3z}$ rotation. On the other hand, the inversion transposes the
two V atoms in the unit cell revealing their equivalence. 
For the I atoms, the $C_{3z}$ operation permutes cyclically 1-3 and 4-6 atoms 
whereas the inversion
transposes pairs of atoms: 1$\leftrightarrow$4, 2$\leftrightarrow$6, 3$\leftrightarrow$5.  
As a result, all I atoms are equivalent.
In general, the magnetic structure can reduce the symmetry of the system 
below the symmetry of the atomic lattice. In VI$_3$, the ferromagnetic structure with
moments parallel to the $z$ axis preserves the symmetry operations of the lattice \cite{comment}.

The symmetry analysis performed above can be used to address the following
important question: Can the $z$ axis 
be the easy axis of the ferromagnetic VI$_3$ with $R\overline{3}$ structure? This question is equivalent 
to the question whether the
ferromagnetic structure with the moments parallel to the $z$ axis
is distinguished by symmetry with respect to the magnetic structures obtained by the 
infinitesimal deviations of the atomic moments from the $z$ direction 
\cite{Sandratskii1995,Sandratskii1998,Sandratskii2020}.
Indeed, if the $z$ axis is not distinguished by symmetry 
it is just one of the continuum of axes that, before the concrete numerical
study is carried out, appear as equally possible
realizations of the easy axis.
The probability that one axis selected by us from the continuum of formally equivalent
axes corresponds to the desired property is negligible.
In this case, the calculation started with the atomic moments oriented parallel 
to the $z$ axis is predicted to result in the deviation of the moments 
from this axis. In the iterative process, the moments tend to assume self-consistent directions.
These directions are accidental in the
sense that they are not characterized by an additional symmetry and cannot be determined 
without direct calculations.

On the other hand, if a selected axis is distinguished by symmetry with respect to 
the axes obtained by infinitesimal deviations, we deal with the symmetry constraint
on the magnetic structure (or, equivalently, symmetry protection of the magnetic structure).
The calculations started with symmetry protected directions of the atomic moments
preserve these directions during iterations. The symmetry protected directions of
the magnetic moments are natural candidates for the magnetic easy axis. 

Applying these general principles to VI$_3$ in $R\overline{3}$ we notice that the $C_{3z}$ symmetry operation
is disturbed by any deviation of the V moments from the $z$ axis. This makes
the $z$ direction of the V moments to be symmetry protected and reveals
the $z$ axis as a symmetry supported option for the easy axis. 
Both spin and OMs of the V atoms must be collinear to the $z$ axis
to satisfy the symmetry constraint.
Therefore, they must be collinear also to each other. 
However, for the moments induced on the I atoms the situation is different.
The symmetry operations transform I atoms into each other and,
therefore, impose constraints not on the directions of the moments of 
individual atoms but on the relative directions of the moments
of different atoms: If a symmetry operation $\alpha$ transforms the position of atom $i$ 
in the position of atom $j$, it also transforms the moment of atom $i$ in the
moment of atom $j$.
Since there is no symmetry operation 'responsible' for preserving the collinearity of the I
moments to the $z$ axis, the symmetry analysis predicts the deviation of the I moments from the $z$ axis.
As there is no symmetry requirement of equal deviations of the 
spin and orbital moments, the spin and orbital moments of the I atoms are expected to deviate 
by different angles and to
become noncollinear to each other. As a general rule, always when the atomic moments
assume accidental directions, the spin and orbital
moments of the same atom are noncollinear.  
The calculations fully confirmed the predictions following from the symmetry analysis.

\subsection{Distorted atomic structure}
\label{Sec_sym_AD}

If we take into account the AD-structural distortion (Fig.~\ref{Fig_lat}b) the symmetry of the lattice decreases.
The $C_{3z}$ symmetry is broken whereas the inversion remains intact.
The breaking of the symmetry with respect to the $C_{3z}$ rotation makes the deviation of the 
V moments from the $z$ axis inevitable. The remaining inversion symmetry 
preserves both the equivalence of the V atoms and the constraint on the moments 
of the two V atoms to be parallel.
This correlates closely with the results of experiments reporting the deviation
of the magnetization direction from the $z$ axis. 
We emphasize that the inversion symmetry imposes the constraint of parallelity separately on
spin and orbital moments but does not require the collinearity of the spin and orbital moments
to each other. 
Therefore the intraatomic noncollinearity of the V spin and orbital moments 
is one of the consequences of the broken $C_{3z}$ symmetry 
(Fig.~\ref{Fig_mod}).  

The effect of the distortion on the I atoms is more complex. The inversion makes equivalent 
the pairs of atoms transposed by this operation (see above). The moments of equivalent atoms are parallel. The
atoms of different pairs are inequivalent and their moments are not transformed into each
other by a symmetry operation. 

If we consider a more complex distortion that breaks also the inversion symmetry,
all atoms become inequivalent. Respectively, there is no symmetry determined 
relations between values and directions of the spin and orbital magnetic moments 
of either the same atom or different atoms. 

\section{Results of calculations}
\label{Sec_Results}

\subsection{DFT calculations}
\label{Sec_GGA}
\begin{figure}[t]
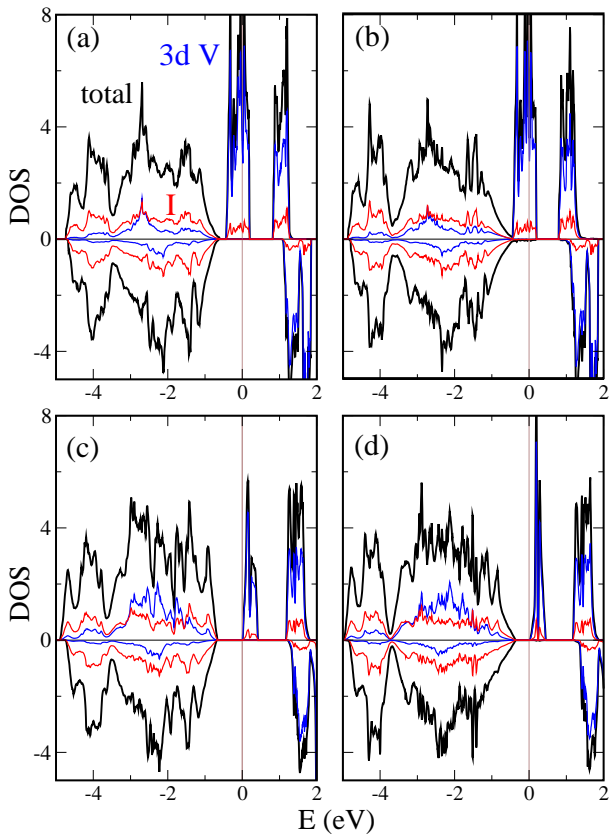

\includegraphics*[width=8cm]{FIG_DOS_IA.eps}
\includegraphics*[width=8cm]{FIG_DOS_IIA.eps}
\caption{Densities of states. Above the abscissa axis are the spin-up DOSs, below the
abscissa axis are the spin-down DOSs. Black lines present total DOSs in numbers o
states per 1~eV and formula unit, blue and red lines present 
partial V 3d and I DOSs per one atom. The energy origin is at the Fermi level in the 
cases of metallic state [panels (a) and (b)] and at the bottom of the valence band
in the insulating states  [panels (c) and (d)].
(a) Calculation without SOC and $U$=0.
(b) Calculation with SOC and $U$=0.  
(c) Calculation without SOC and $U$=2.72~eV.  
(d) Calculation with SOC and $U$=2.72~eV.  
}
\label{Fig_DOS}
\end{figure}

In the calculations for the $R\overline{3}$ lattice we used experimental lattice parameters
$a$=6.835~{\AA}, $c$=19.696~{\AA}\cite{Tian2019}. 
In Fig.~\ref{Fig_DOS}a we show the density of states (DOS) calculated without 
account for the SOC and electron correlations.
The characteristic feature of the
electronic structure is its half-metallic nature: the spin-up subsystem is metallic and spin-down
subsystem is insulating. The spin moment is exactly 2~$\mu_B$ per formula unit (FU). 
The value of the V spin moment is 2.124~$\mu_B$. The OM is
quenched.

There is a group of the spin-up V 3d states around the Fermi level 
that plays important role in the discussion of the results of the DFT+$U$ calculations presented below. 
This group of states
is associated with $t_{2g}$ orbitals originating in the crystal-field splitting of the V 3d states in the 
octahedral environment of the I ligands. 
To recall, the splitting of the 3d states to $t_{2g}$ and $e_g$ subsystems appears 
in the standard consideration of the influence of the octahedral ligand environment 
on a 3d atom. 
In the case of VI$_3$ crystal the octahedron axes are 
different from the $z$ axis and are not symmetry axes of the crystal. 
Although the models based on the treatment of the crystal-field splitting of the V 3d states 
in the octahedral environment are useful for qualitative considerations 
they are a substantial simplification of the real physical situation. 
The DFT-based calculations reflect more closely the actual complexity of the system.
The results of these calculations will be the basis of our considerations. 
Keeping this comment in mind, we for brevity will refer to the spin-up DOS peak around the 
Fermi energy as $t_{2g}$-peak.   

The account for the SOC does not change the general structure of the DOS (Fig.~\ref{Fig_DOS}b) leading, 
however, to important differences in the details. Strictly speaking, the DOS is not now half-metallic
since the SOC mixes spin-up and spin-down states and, therefore,
there is no 100\% spin-polarization of the electronic states at the Fermi level. 
On the other hand, the contribution of the spin-down states at the Fermi level is 
small and hardly noticeable in Fig.~\ref{Fig_DOS}b. The spin moment of V is 2.145~$\mu_B$.
The V OM is unquenched and assumes the value of  $-$0.075~$\mu_B$
where negative sign means the direction opposite to the direction of the spin moment.
The spin and orbital moments of the V atoms are collinear to the $z$ axis.

\begin{figure}[t]
\includegraphics*[width=8cm]{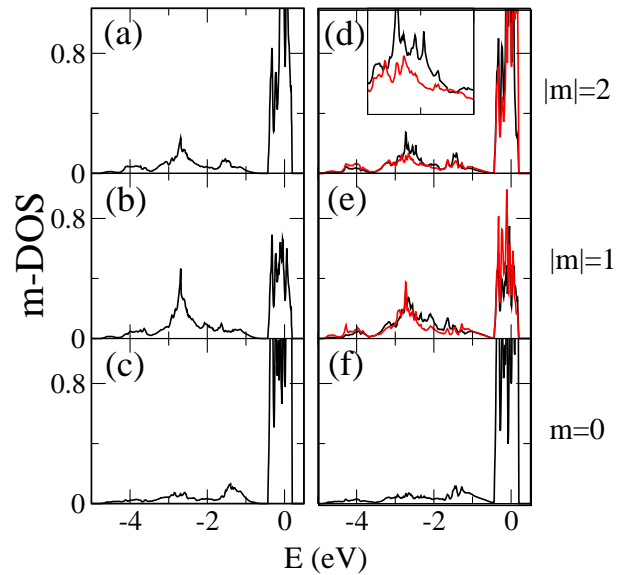}
\caption{Partial V 3d spin-up $m$-DOSs. (a)-(c) calculations without SOC and U=0;
(d)-(f) calculations with SOC and $U$=0. 
(a),(d) $m=-2,2$; (b),(e) $m=-1,1$; (c),(f) $m=0$. 
In the calculation without SOC, DOS$_m$=DOS$_{-m}$.
In panels (d),(e) $m$-DOS for negative $m$ are in red.
\textcolor{black}{
The insert in panel (d) illustrates on a larger scale the difference 
between $m$-DOSs with  $m=-2$ and  $m=2$ in the energy interval 
between -3 and -2 eV.
}
}
\label{Fig_m-DOS_U00}
\end{figure}
To illustrate the origin of the nonzero V OM we  
show in Fig.~\ref{Fig_m-DOS_U00} the $m$-resolved V 3d DOSs. 
\textcolor{black}{
The partial DOS$_m$ corresponding to a given $m$ is obtained by projecting of the electron states 
to complex spherical harmonic $Y_{2m}$. The sum of the $m$-resolved DOSs gives the total
V 3d DOS.
}
The complex spherical harmonics are defined with 
respect to the $z$ axis. 
\textcolor{black}{
The nonzero atomic OM is the result of different occupation of the $m$ and $-m$ states
[see also Eq.~(\ref{Eq_m_oz})].
}
Without SOC ( Figs.~\ref{Fig_m-DOS_U00}a-c),  DOS$_m$=DOS$_{-m}$ and the $z$ projection 
of the OM is zero. With SOC taken into account (Figs.~\ref{Fig_m-DOS_U00}d-f),  
DOS$_m$$\neq$DOS$_{-m}$ and 
the $z$ projection of the OM is unquenched.

In Sec.~\ref{Sec_GGA+U} we will discuss the results of the DFT+$U$ calculations 
where the orbital density matrix $n$ enters the secular matrix of the method and
influences directly the formation of the electronic structure. 
Since the symmetry 
determined properties of the occupation matrix are identical in both DFT and DFT+$U$ calculations
it is useful to consider these properties already here.
The calculations give the following form of the $n$ matrix 
\begin{equation}
\left(\begin{array}{ccccc} 
 * &   0 &   0&   * &   0\\  
 0 &   * &   0&   0 &   *\\ 
 0 &   0 &   *&   0 &   0\\ 
 * &   0 &   0&   * &   0\\ 
 0 &   * &   0&   0 &   *
\end{array}\right)
\label{Eq_n_matrix}
\end{equation}
Equation~\ref{Eq_n_matrix} shows a spin-diagonal block of matrix $n$.
The rows and columns of the matrix are numbered with the magnetic 
quantum number $m=2,1,0,-1,-2$. 
The asterisks, $*$, present nonzero elements of the matrix: 
the diagonal elements $n_{m,m}$,
and four off-diagonal elements $n_{2,-1}$, $n_{-1,2}$, $n_{-2,1}$, $n_{1,-2}$. 

The nonzero off-diagonal elements of $n$ reveal that the symmetry supports the hybridization of
the pairs of the atomic orbitals: 
orbital $m$=$2$ with orbital $m$=$-1$ and
orbital $m$=$1$ with orbital $m$=$-2$. The origin of this hybridization can be
explained as follows. The symmetry operations of the rotation about the $z$ axis
by angles 0$^\circ$, 120$^\circ$ and 240$^\circ$ form an abelian group with three one-dimensional
irreducible representations:
$\Gamma_1(1,1,1)$, $\Gamma_2(1,\epsilon,\epsilon^{-1})$, $\Gamma_3(1,\epsilon^{-1},\epsilon)$ 
where $\epsilon$=$\exp(i2\pi/3)$ and the numbers in the parentheses correspond to the three symmetry operations.
The $m$=$0$ spherical harmonic transforms according to $\Gamma_1$,  $m$=$2$ and $m$=$-1$ harmonics transform
according to $\Gamma_2$  and  $m$=$1$ and $m$=$-2$ harmonics transform according to $\Gamma_3$. 
The non-zero off-diagonal elements of $n$ reflect the hybridization between orbitals belonging 
to the same irreducible representation. 

The elements of the $n$ matrix determine the value of the OM.
The $z$ component of the OM, $m_{oz}$, is determined by the diagonal elements
\begin{equation}
m_{oz}=\sum_m m \: n_{m,m},
\label{Eq_m_oz}
\end{equation} 
whereas the $x$ and $y$ 
components, $m_{ox}$ and $m_{oy}$, depend on the values of the off-diagonal elements
\begin{equation}
m_{ox} \pm i m_{oy} = \sum_m  {(2 \mp m)[3 \pm (m-1)]}^\frac{1}{2} n_{m,m\mp1}.
\end{equation}  
In general, nonzero off-diagonal elements can lead to nonzero $x$ and $y$ components of the OM.
However, this needs nonzero elements $n_{mm'}$ with $|m-m'|$=1. In our case, these matrix elements are zero that 
leads to vanishing $x$ and $y$ components of the OM. Therefore, the form of the calculated $n$ matrix is consistent with 
the conclusion about the collinearity of the V OM to the $z$ axis made above. 

For the magnetic moments of the I atoms the calculations with account for 
SOC gave the following results.
The spin moments have the value of 0.023~$\mu_B$ 
and deviate from the negative direction of the $z$ axis by 0.5$^\circ$. 
The OMs have the value of  0.022~$\mu_B$
and deviate from the negative direction of the $z$ axis by 7.3$^\circ$.
The in-plane projections of the I moments compensate each other. 
The directions of both spin and OMs are close to the negative direction
of the $z$ axis. This feature 
corresponds to the expectation based of the third Hund's rule:
for an isolated I atom the spin and OMs are parallel to each other. On the other
hand, the sizable deviations of the moments from the $z$ axis manifest the influence
of the lattice. The fact that the induced I spin moments 
deviate much weaker from the $z$ axis 
than the induced I OMs reflects a stronger influence of the atomic lattice on the 
orbital degrees of freedom. 

\subsection{DFT+$U$ calculations}
\label{Sec_GGA+U}

Although our DFT calculations discussed in Sec.~\ref{Sec_GGA}
provide some insight into the formation of the properties of the 
system, they result in the metallic state contradicting the 
experiments revealing the formation of an insulating state. As most of the previous theoretical
studies, to deal with the Mott insulating state of VI$_3$ we made use of the DFT+$U$ method.

Indeed, the calculations result in the splitting of the $t_{2g}$ DOS peak and the formation of 
the insulating gap between occupied and empty states. 
A remarkable result of our DFT+$U$ calculations
is the possibility to obtain self-consistent insulating states of VI$_3$ with strongly different character 
of the splitting of the $t_{2g}$ peak.
This possibility of different self-consistent results of the DFT+$U$ calculations is in correlation
with the conclusions of Wang {\it et al.} \cite{Wang2020}. We, however, emphasize that our 
calculations did not give a metallic state of VI$_3$ as
reported in Ref.~\onlinecite{Wang2020}. Importantly, the different insulating states obtained in our calculations 
are connected with the formation of different OMs. 
In this respect our results agree with the physical picture suggested by Yang {\it et al.} \cite{Yang2020}. 
It is also worth noting that in Ref.~\onlinecite{Huang2020} the possibility of two different orbital ordered
states is discussed where both insulating states have zero OM.

\begin{figure}[t]
\includegraphics*[width=8cm]{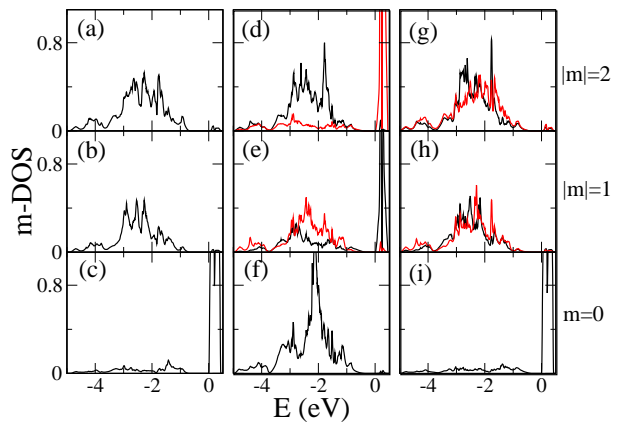}
\caption{Partial V 3d spin-up $m$-DOSs calculated with $U$=2.72~eV. (a)-(c) calculations without SOC;
(d)-(f) calculations with SOC, large OM;
(g)-(i) calculations with SOC, small OM. 
(a),(d),(g) $m=-2,2$; (b),(e),(h) $m=-1,1$; (c),(f),(i) $m=0$. 
In the calculation without SOC, DOS$_m$=DOS$_{-m}$.
In panels (d),(e),(g),(h) $m$-DOS for negative $m$ are in red.
}
\label{Fig_m-DOS_U}
\end{figure}
In Fig.~\ref{Fig_DOS}c we show the DOS obtained with $U$=2.72~eV and with SOC being neglected. 
At the beginning of the iterations the $n$ matrix was set to zero. In the absence of the SOC
the OM remained quenched in the iterations that is reflected
in the property that the partial DOS corresponding to magnetic quantum number $m$ 
is identical to the partial DOS corresponding to $-m$ (Fig.~\ref{Fig_m-DOS_U}ab).
This degeneracy combined with the inter-orbital hybridization, $m$=$-2$ with $m$=$1$ 
and $m$=$2$ with $m$=$-1$ (see Sec.~\ref{Sec_GGA}),
leads to the joint shift to lower energies of the occupied parts of the corresponding four partial DOS. 
These states are separated by a gap from the empty peak of predominantly $m$=$0$ states (Fig.~\ref{Fig_m-DOS_U}c). 

On the other hand, if we include the SOC 
already the first iteration results in the different occupation of the $m$ and $-m$ orbitals
unquenching the OM.
This orbital polarization is strongly enhanced in the course of the DFT+$U$ iterations leading
to the splitting of the occupied and empty states (Figs.~\ref{Fig_DOS}c,\ref{Fig_m-DOS_U}d-f) 
and large orbital 
moment of $-$0.782~$\mu_B$. The occupied
partial DOSs correspond now mainly to the hybridized $m$=$-2$ and $m$=$1$ orbitals and to the $m$=$0$ orbital
 (Fig.~\ref{Fig_m-DOS_U}d-f). 

We obtained also the self-consistent state with distinctly smaller V OM of  0.111~$\mu_B$.
This calculation was performed as follows. First the SOC was switched on only on the 
V atoms. We obtained the V OM of 0.093~$\mu_B$. Then the SOC was switched on
also on the I atoms. 
In this case, the structure of the $m$-DOSs and, therefore, the nature of the 
splitting of $t_{2g}$ peak (Fig.~\ref{Fig_m-DOS_U}g-i) is rather close to the case 
with quenched OM (Fig.~\ref{Fig_m-DOS_U}d-f).     
The energy of the state with small OM is $\sim$34 meV/FU higher than the energy of the state with large OM.

The possibility to obtain
insulating states with different OMs can be interpreted as follows. 
There are two different aspects to mention. On one hand, the hybridization with the environment of 
various $m$-orbitals of the electron states belonging to the $t_{2g}$ DOS peak 
is not strong enough to counteract efficiently 
the splitting tendencies initiated by the on-site correlation governed by the $U$ parameter.
Therefore the insulating gap can appear between states with different types of atomic orbitals.
The types of the states that become stronger occupied and, therefore, lower their energy leading to the
insulating gap determine the value of the 
OM formed by the occupied states.   
On the other hand, the interatomic hybridization is also an important factor
that cannot be neglected.
This, in particular, is reflected in the closely correlated behavior of the pairs of the 
partial $m$-DOSs: $m$=$-2,1$ and $m$=$2,-1$. In an isolated atom, the states corresponding
to different values of $m$ do not hybridize. 

Although the account for on-site correlations changes the values
of the atomic moments, the symmetry properties remain intact. As the result,
the V spin and orbital moments are collinear to the $z$ axis. 
Since, however, the V spin moments changes only weakly while the orbital moment
increases by more than 10~times, the total V moment decreases strongly 
to the value of 1.362~$\mu_B$. 
The large value of the V OM is in 
agreement with the physical picture suggested in Ref.~\onlinecite{Yang2020} and with the reduced magnetization
values reported in a number of experimental works \cite{Son2019,Kong2019,Liu2020}.

Examining the calculated I moments, we find that the I spin moments deviate from 
the negative direction of the $z$ axis by 4.3$^\circ$ whereas the deviation of the orbital
moments reaches 27.5$^\circ$. We conclude that the correlation of the V 3d 
electrons is transferred by means of interatomic hybridization to the I states leading 
to strong enhancement of the deviation of the I moments from the $z$ axis and, as a result,
to strong enhancement of the intraatomic  noncollinearity of the spin and orbital moments. 

\subsection{Distorted atomic structure}

\subsubsection{AD-Distortion with preserved inversion symmetry}
\label{Sec_Di1}

As discussed in Sec.~\ref{Sec_sym_AD}, the AD-distortion reduces the lattice symmetry. 
Since the $z$ axis is not a symmetry axis of the system 
it cannot be the easy axis and the V moments are predicted to deviate from the $z$ axis. On the other hand, 
the inversion symmetry remains intact that preserves the equivalence of the V atoms
and the collinearity of the moments of the V atoms. 
The direction of the moments is 
'accidental' in the sense that it cannot be predicted on the symmetry basis. 
Therefore, although the V spin moments are parallel to each other and
the V orbital moments are parallel to each other, the spin moments are noncollinear to the orbital moments.
Also the consequences of the AD-distortion for the symmetry properties of the I atoms
are discussed in Sec.~\ref{Sec_sym_AD}.

The calculations 
for the AD-distorted structure \cite{AD_comment}
confirm the predictions of the symmetry analysis. If we start with the 
V moments parallel to the $z$ axis, after the first iteration
the moments deviate from the $z$ direction 
reflecting the breaking of the symmetry protection of the collinearity to the $z$ axis.
The magnetic moments of different V atoms remain parallel to each other. However,
the spin and orbital moments of the same atom become noncollinear.
As expected, the form of the $n$ matrix changes compared with Eq.~(\ref{Eq_n_matrix}): 
the off-diagonal matrix elements responsible
for the $x$ and $y$ components of the OM are now nonzero.  

\begin{figure}[t]
\includegraphics*[width=8cm]{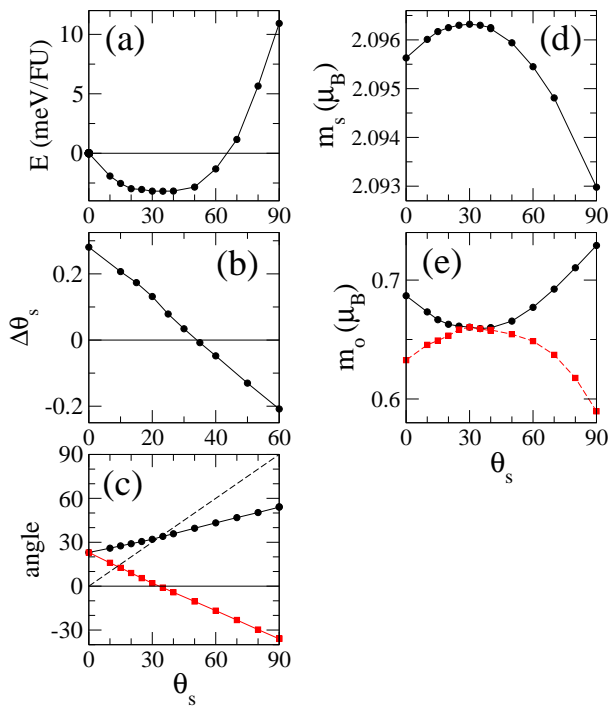}
\caption{Calculated physical quantities as functions of the constrained spin angle $\theta_s$
of the V atoms. All angles are in degrees. 
(a) the energy of the system, 
(b) the difference between the constrained angle $\theta_s$ and the spin angle $\theta_s'$ 
at the end of the iteration,
(c) the direction of the orbital moment $\theta_o$ (black line, filled circles) and 
difference $\theta_o-\theta_s$ (red line, filled squares),
(d) the value of the spin moment,
(e) the value of the orbital moment (black line, filled circles) and the projection
of the orbital moment on the constrained direction of the spin moment (red line, filled squares).
}
\label{Fig_e_of_theta}
\end{figure}
To determine the easy axis direction we performed the following set of calculations. 
We constrained several directions of the 
V spin moments in the $xz$ plane. For each of these directions specified,
by angle $\theta_s$,  
other quantities were allowed to relax towards self-consistency. The total energy of the system
as the function of $\theta_s$ is presented in Fig.~\ref{Fig_e_of_theta}a. We see that at $\theta_s$=0 
the slope of the curve 
is nonzero and, therefore, this sate does not correspond to an extremum of energy. This feature confirms
the results of the symmetry analysis. With increasing $\theta_s$, the energy first decreases. 
Then it becomes rather flat reaching a flat minimum at $\sim$35$^\circ$. For $\theta_s$ 
above $\sim$60$^\circ$ the energy quickly
increases revealing the presence of strong magnetic anisotropy in the system. 

At each iteration the calculated direction of the V spin moment given by angle $\theta_s'$ 
is not equal to the initial direction given by $\theta_s$.
Due to constraint, this difference is neglected and the initial value $\theta_s$ is preserved in the iterations. 
In Fig.~\ref{Fig_e_of_theta}b we plot the difference $\theta_s'-\theta_s$ as a function of $\theta_s$.
This difference becomes zero at $\sim$35$^\circ$ showing that the position of the 
energy minimum corresponds to the unconstrained self-consistent state.

Another point of interest is the properties of the orbital moment. As stated above
on the basis of the symmetry arguments,
the spin and orbital moments of the V atoms in the distorted lattice become noncollinear.
We denote as $\theta_o$ the angle of the deviation of the calculated orbital moment
from the negative direction of the $z$ axis.
In Fig.~\ref{Fig_e_of_theta}c we present $\theta_o$ and $\theta_o$-$\theta_s$ as functions of $\theta_s$.
We obtain remarkably strong noncollinearity of the two
V moments. The angle between the moments is 23$^\circ$ for $\theta_s$=0$^\circ$. It 
decreases almost linearly with the increase of $\theta_s$. At $\theta_s$$\sim$35$^\circ$ two moments
become collinear. With further increase of $\theta_s$ the noncollinearity increases
again. At $\theta_s$=90$^\circ$ the angle between spin and orbital moments reaches 35$^\circ$.

A number of important conclusions can be derived from these graphs. First,
the interval of the variation of the orbital moment direction is distinctly 
smaller than the interval of the variation of $\theta_s$. This reveals a limited
influence of the spin moment direction on the direction of the orbital moment.
The competing stronger influence is exerted by the lattice. Second, in the self-consisted
state at $\theta_s$$\sim$35$^\circ$ the two atomic moments are almost collinear. 
The angle of $\sim$35$^\circ$ specifying the direction of the easy axis is in very good 
correlation with the experimental measurements \cite{Yan2019,Dolezal2019}.

It is also interesting to inspect the magnitudes of the spin and orbital atomic
moments as the functions of $\theta_s$. The corresponding data are plotted in
Figs.~\ref{Fig_e_of_theta}de. The variation range of the spin moment value of the is distinctly
smaller than the variation range of the orbital moment. A flat
maximum of the spin moment is reached at $\sim$35$^\circ$, that is very close to the 
direction of the easy axis. In the same region the magnitude of the orbital
moment assumes its minimum (Fig.~\ref{Fig_e_of_theta}e). The minimum of $m_o$ close to the 
the easy axis direction appears surprising
taking into account that the widely accepted analysis of Bruno \cite{Bruno1995} predicts the maximal value
of the OM for the 
easy axis direction. However, if we take into account that the direction of the 
OM deviates from the direction specified by $\theta_s$ and consider 
the value of the projection of vector {\bf m}$_o$ on the constrained direction of the 
spin moment we obtain the curve
characterized by a flat maximum at $\sim$35$^\circ$ (Fig.~\ref{Fig_e_of_theta}e). The behavior of the projection
of the OM on the constrained direction of the spin moment correlates with the 
Bruno's conclusion. Nevertheless, it is worth emphasizing that 
the Bruno's treatment deals with elemental 3d ferromagnets in a highly
symmetric atomic lattice whereas we investigate a compound with
low symmetry lattice and strong SOC on ligands. Therefore, more complex
relationships between physical quantities should be expected.                           

For completeness we include some information on the properties of the I moments.
First, we will take the case of $\theta_s$=0$^\circ$. 
As predicted by the symmetry analysis, the I atoms become inequivalent. The I spin moments
of atoms 1-3 have the values 0.040, 0.013, 0.014~$\mu_B$
and deviate from the $z$ axis by 1$^\circ$, 12$^\circ$ and 9$^\circ$, respectively. The OMs 
have the values 0.022, 0.034 and 0.033~$\mu_B$ deviating by 18$^\circ$, 34$^\circ$ and 34$^\circ$ 
from the $z$ axis. As a result we have a
complex magnetic configuration with strong intraatomic noncollinearity of the spin and 
orbital moments and strong noncollinearity of the inducing and induced moments.   
For the self-consistent state corresponding to $\theta_s$=35$^\circ$ we obtained the values
of the I orbital moments 0.024, 0.036, and 0.035~$\mu_B$ that are very close to the values
for $\theta_s$=0$^\circ$. The noncollinearity with 
respect to the V moments is given by 2$^\circ$, 10$^\circ$, and 10$^\circ$ that is now distinctly smaller than
in the case of $\theta_s=0^\circ$.

\subsubsection{AD-Distortion combined with broken inversion symmetry}
\label{Sec_Di1+}

As discussed in Sec.~\ref{Sec_Di1}, the preserved inversion symmetry of the AD-distorted atomic structure is
responsible for the equivalence of the V atoms and collinearity of their magnetic
moments. Consequently, the breaking of the inversion symmetry must lead to the inequivalence 
of the V atoms and noncollinearity of their V moments. In this work we will consider the distortion 
of the I environment of the V atoms simulated by the shift of all V atoms along the $y$ axis 
keeping the positions of the I unchanged. This shift leads to the breaking of the inversion 
symmetry. These calculations have a model character and are not based on the 
experimental information on this type of the atomic lattice distortion.

We considered three values of the shift $\delta_1$=0.034~{\AA}, $\delta_2$=2$\delta_1$, $\delta_3$=4$\delta_1$.
For all three cases we performed calculations with the directions of the V spin moments 
constrained parallel to the $z$ axis. 
As expected, the V atoms in this type of lattice are inequivalent. 
For the smallest shift $\delta_1$ we obtained very small
difference in the value of the spin moments of the V atoms: $m_s$(V1)=2.092,
$m_s$(V2)=2.086~$\mu_B$.   
However, for the orbital moments the difference is considerable: the values of the moments
are 0.735 and 0.608~$\mu_B$, the deviations from the $z$ axis are ~20$^\circ$ and ~28$^\circ$. 
For shift $\delta_2$ we obtained large difference between the values of the spin moments
$m_s$(V1)=2.020  $m_s$(V2)=1.355~$\mu_B$.
For the orbital moments we get the values  0.667, 0.688~$\mu_B$ and 
the deviation angles 13$^\circ$ and 24$^\circ$.
For the largest shift $\delta_3$, the spin moments are 2.064 and 0.035~$\mu_B$.
The values and deviations of the orbital moments are 0.628 and  0.015~$\mu_B$ and 26$^\circ$ and 141$^\circ$.

Analysis of these data shows that distortion of the I environment of the V atoms 
has strong influence on the magnetic moments of the V atoms. This influence is
complex and nonlinear with respect to the value of the distortion. For the smallest
distortion $\delta_1$ the spin moments of the two V atoms remain almost equal whereas the 
values and directions of the orbital moments differ considerably. Doubling the 
shift we obtain strong drop of $m_s$(V2) which looses a half of its value. Surprisingly,
here the values of the orbital moments are rather close to each other. For the largest shift $\delta_3$
we obtain dramatic change of the magnetism of the V2 which is now nearly nonmagnetic.

The model calculations discussed in this section show (i) the distortion of the I environment
of the V atoms has a strong effect on the magnetic characteristics of the V atoms. (ii) As predicted
the inversion symmetry breaking leads to the inequivalence of the V atoms.
This feature correlates with the experimental results of Gati {\it et al.} \cite{Gati2019}.
(iii) The character of the changes is strongly non-linear with respect to the strength of the distortion.
This means that quantitative conclusions about the properties of the system need precise information 
on the character of the lattice distortions. 

\section{Conclusions}
\label{Sec_conclus}

Recently discovered ferromagnetism of the layered vdW material VI$_3$ attracts much research attention, both theoretical
and experimental. Despite substantial progress, in some important aspects no consensus
has been reached. Among the experimental questions where no consensus is achieved are 
(i) the deviation of the easy axis from
the normal to the VI$_3$ layers, (ii) a possible inequivalence of the V atoms, (iii) the value of the V 
magnetic moments. The theoretical works differ in the conclusions on the conduction nature of the system, 
the value and the role of the V orbital moments. To the best of our knowledge there is no 
theoretical works addressing issues (i) and (ii) and only one work dealing with the reduced
value of the V moment and explaining it by the formation of a large orbital moment.

By combining the symmetry arguments with DFT and DFT+$U$ calculations we have shown
that the distortion of the crystal structure suggested by Son {\it et al.} \cite{Son2019} must lead to the deviation of the 
easy axis from the normal to the VI$_3$ layers in close correlations with the experimental
results reported in Ref.~\onlinecite{Yan2019,Dolezal2019}. The AD accompanied by the distortion of the 
I environment of the V atoms
leads to the breaking of the inversion symmetry and inequivalence of the V atoms.

\textcolor{black}{
In agreement with earlier works \cite{An2019,Yang2020,Wang2020} our calculation show that  
the DFT+$U$ method provides
a proper account for the interplay between interatomic hybridization and on-site electron
correlations leading to the formation of the Mott insulating state. The standard DFT calculations
give a metallic state of VI$_3$ contradicting the experimental results. 
}

Our DFT+$U$ calculations result in large values of the OMs of the V atoms
leading to reduced total V moment, in agreement with a number of experimental results
and with the physical picture suggested by Yang {\it et al}. \cite{Yang2020}. We obtained large 
intraatomic noncollinearity of the V spin and orbital moments revealing strong competition   
of the effects coursed by the on-site electron correlation, SOC, interatomic hybridization.

Our calculations confirm the experimental result of strong magneto-elastic 
interaction revealing itself in the strong dependence of the magnetic properties 
on the distortion of the atomic structure. This makes the account for the lattice distortion
crucial for the understanding
of the magnetic properties of VI$_3$. 

Our study agrees with previous reports about possibility of the multiple convergence of the 
DFT+$U$ method. However, in contrast to Ref.~\onlinecite{Huang2020} in our calculations all converged 
states are Mott insulators.
The different states are characterized by different values of the OM with the 
state with the large OM being the lowest in energy. 

We hope that our study contributes importantly to the understanding of the
magnetic properties of VI$_3$ and will 
stimulate further experimental and theoretical investigations of VI$_3$ and other 
vdW and 2D magnets. 

\section{Acknowledgments}
One of the authors  (KC) acknowledges financial support of the Czech Science Foundation, 
project No. 19-16389J.

\end{document}